\documentclass[12pt]{article}

\usepackage{graphicx}
\usepackage{appendix}

\usepackage{amssymb}	% Allows use of AMS's mathematical symbols
\usepackage{amsmath}    % Allows use of AMS's mathematical commands
\usepackage{algorithmic}
\allowdisplaybreaks

\newcommand{\qed}{\nobreak \ifvmode \relax \else \ifdim\lastskip<1.5em \hskip-\lastskip \hskip1.5em plus0em minus0.5em \fi \nobreak \vrule height0.75em width0.5em depth0.25em\fi} 
\def\0{\bf \0}

\def\0{{\bf 0}}

\def\R{\mathbb{R}}

\def\g{{\bf g}}
\def\h{{\bf h}}

\def\x{{\bf x}}

\usepackage{color}
\usepackage[normalem]{ulem}

\begin{document}

%\maketitle

%\title{An infeasible interior-point arc-search algorithm for nonlinear constrained optimization}
\title{On optimal LISA orbit design}

\author{Yaguang Yang\thanks{NASA, Goddard Space Flight Center, 8800 Greenbelt Rd, Greenbelt, 20771 MD \\
Email: yaguang.yang@nasa.gov}}

%\affiliation{NASA, Goddard Space Flight Center, 8800 Greenbelt Rd, Greenbelt, 20771 MD}

%\date{Received: date / Accepted: date}
% The correct dates will be entered by the editor

\maketitle

\begin{abstract}

The ESA/NASA joint LISA (laser interferometer space antenna) mission is designed to 
detect gravitational waves, which relies crucially on maintaining three-spacecraft 
constellation as close to an equilateral triangle with a designed distance
as possible. Efforts have been made to achieve this goal by using various simplified 
models to make it easy to approximately solve the complex problem. In this paper, the 
problem is formulated as a nonlinear optimization problem using exact nonlinear Kepler's 
orbit equations. It is shown that the optimal solution based on the exact nonlinear Kepler's 
orbit equations gives a better solution than the previously obtained ones.

\end{abstract}

%\vspace{0.1in}
%\noindent
%{MSC class:  }

\vspace{0.2in}
\noindent
{\bf keywords:} LISA; formation fly; optimal orbit design

\section{Introduction}

LISA is an ESA/NASA mission with the objective of sensing low-frequency gravitational waves. 
Two sets of interferometers are installed on each of the three spacecraft. Each spacecraft's
orbit is on a heliocentric plane.  The three spacecraft are the three vertices of a moving 
(approximate) equilateral triangle with time-dependent distance variations around a
designed length \cite{dnkv05}. This variation introduces Doppler shifts and breathing 
angles which affect the interferometers' measurement \cite{mpb12,nkdv06}. It is desirable 
to minimize the variation to meet the constraints of the motion range of the Optical Assembly 
Tracking Mechanism and the bandwidth limitation of the LISA phasemeter. The design of the 
best orbits that minimizes the distance variations about a designed length between the
three spacecraft is a difficult nonlinear dynamics problem, therefore, various approximations 
have been considered by previous authors so that simplified problems based on approximate 
models can be solved \cite{amato19,dnkv05,hughes02,mpb12,nkdv06}. In \cite{hughes02}, 
the author searched for a solution using all six Keplerian elements. In \cite{mpb12}, the 
Clohessy–Wiltshire (a first-order) model \cite{cw60} is used to reduce the 
deviation in distance between the three spacecraft. In \cite{dnkv05}, a first-order 
approximation model is used and shows that the three-spacecraft constellation flies 
in an approximate equilateral triangle formation (we will refer to this orbit as the
DNKV orbit). In \cite{nkdv06}, a second-order approximation model is used, 
aimed at improving the solution of \cite{dnkv05} (we will refer to this orbit as the
NKDV orbit).  The orbit obtained in \cite{nkdv06} is used as the baseline\footnote{
This solution can be used as an initial guess for optimizing a full numerical 
nonlinear model which includes multiple gravitating bodies in the solar system \cite{mj21}. 
As for nonlinear optimization problem 
(which may have many local optimizers), a good initial guess is very important.}
for the LISA project \cite{mj21}. Among six Keplerian elements, only the eccentricity 
and the inclination affect the magnitude of the distance variation, we may
consider only these two Keplerian parameters to simplify the problem without sacrificing 
any accuracy in finding the optimal solution. On the other hand, using the exact nonlinear
Keplerian orbit equations rather than an approximated model and optimizing
for two variables (the eccentricity and the inclination) 
rather than just one (the tilt angle) should give us an exact optimal solution. 
In this article, we provide such an optimal solution and compare it with previously
obtained ones. For this reason, we make the same assumptions used in 
\cite{amato19,dnkv05,hughes02,nkdv06} in the discussion that the Sun is the 
only gravitating body in the solar system. 

\section{Two important orbit designs}

\begin{figure}[ht!]
\centering
\includegraphics[width=5in,height=5in]{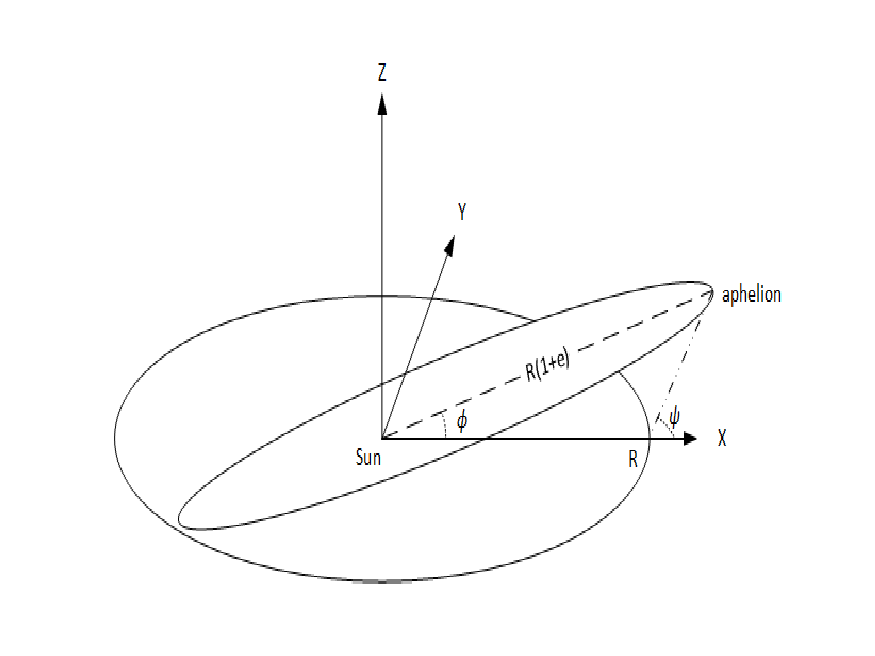}
\caption{The geometry of the LISA spacecraft orbit with respect to the ecliptic plane.} 
\label{orbitGeneral}
\end{figure}

Let the barycentric frame with coordinates (X,Y,Z) be defined as follows: the ecliptic plane
is the X–Y plane and a circular reference orbit on the plane with radius $R=1$ AU is centered 
at the Sun. For an elliptical orbit with its true focus at the Sun (true focus is defined in \cite{wie98}),
let $a$ be the semi-major axis and $e$ be the eccentricity, and $E$ be the 
eccentric anomaly. Following the definition of \cite{dnkv05}, we choose $t=0$ at the time
when the spacecraft is at aphelion, which is above the $+X$ axis.
This coordinate system is different from many books and this choice 
of initial condition results in a positive sign instead of a negative sign in several 
equations below. The geometry of the orbit is shown in Figure 1. 
According to \cite[page 92]{vallado01}, the following relations hold.
\begin{equation}
X=a (\cos(E) + e), \hspace{0.1in} Y=a \sqrt{1-e^2} \sin(E).
\end{equation}
Let $\mu$ be the heliocentric gravitational constant and
$\Omega = \sqrt{\mu/a^3}$ be the mean motion, then Kepler's equation is given by
\cite[page 53]{vallado01}
\begin{equation}
E + e \sin(E) = \Omega t,
\label{kepler}
\end{equation}
where $\Omega t$ is the mean anomaly of the Sun. Let the orbit of 
spacecraft 1 be obtained by rotating the elliptical orbit about the $-Y$ axis 
by $i$ degrees so that its highest point (maximum $Z$) at t = 0 is in the positive
$X$ direction, i.e., at this point, $E=0$, $X=a(1+e) \cos(i)$, $Y=0$, and
$Z=a(1+e) \sin(i)$. In general, the orbit is given by
\begin{subequations}
\begin{gather}
X_1(E)=a (\cos(E) + e) \cos(i), \\ 
Y_1(E)=a \sqrt{1-e^2} \sin(E), \\ 
Z_1(E)=a (\cos(E) + e) \sin(i)
\end{gather}
\label{orbit1}
\end{subequations}
where $E$ is implicitly related to the time $t$ given by (\ref{kepler}). Let the orbit of 
spacecraft 2 be obtained by rotating the orbit of spacecraft 1 about the $Z$ axis by 
$\frac{2}{3} \pi$ and the orbit of spacecraft 3 be obtained by rotating the orbit of 
spacecraft 1 about the $Z$ axis by $\frac{4}{3} \pi$. Then, their orbits can be represented by
\begin{subequations}
\begin{gather}
X_k=X_1(E_k) \cos\left(\frac{2\pi}{3} (k-1) \right)- Y_1(E_k) \sin \left( \frac{2\pi}{3} (k-1)  \right), \\
Y_k=X_1(E_k) \sin\left(\frac{2\pi}{3} (k-1) \right) + Y_1(E_k) \cos \left( \frac{2\pi}{3} (k-1)  \right), \\
Z_k=Z_1(E_k),
\end{gather}
\label{orbits}
\end{subequations}
where $k=1,2,3$, and $E_k$ implicitly depends on $t$ defined by Kepler's equations
\begin{equation}
E_k + e \sin(E_k) = \Omega t - (k-1)\frac{2\pi}{3}.
\label{Ek}
\end{equation}
We will use this coordinate system throughout the paper. Let $\ell=2,500,000$ 
km be the desired distance between any two spacecraft in the constellation, 
also denote $\alpha=\ell/2R$. In the rest of the paper, we will first review two 
existing orbit designs, then propose an optimal orbit design and compare their 
deviations from a desired distance.

\subsection{DNKV orbit}

\begin{figure}[ht!]
\centering
\includegraphics[width=5in,height=4.7in]{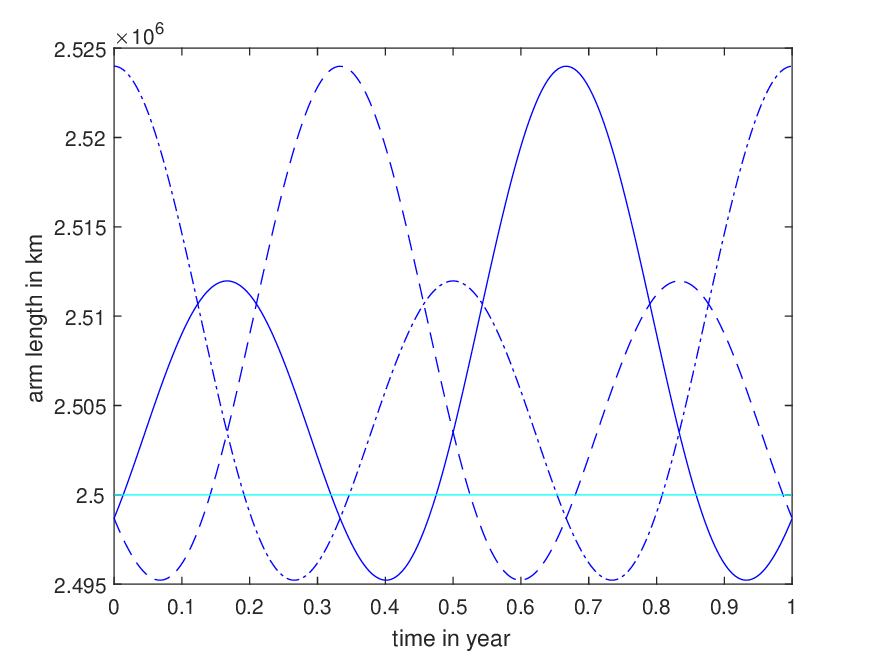}
\caption{Spacecraft distances in one period of DNKV orbit. The horizontal line is the desired distance
between any spacecraft pair. The solid line is the distance variation between spacecraft 1 \& 2.
The dashed line is the distance variation between spacecraft 1 \& 3. The dashed-dot line is the 
distance variation between spacecraft 2 \& 3.} 
\label{orbitDNKV}
\end{figure}

In view of Figure \ref{orbitGeneral}, consider a line segment between the aphelion of 
spacecraft 1 and the intersection of the $X$ axis and the reference circular orbit on the ecliptic plan, 
the angle between the segment and the $X$ axis is denoted as $\psi$ and set 
$\psi=\pi/3$, Dhurandhar et al \cite{dnkv05} have shown
\begin{equation}
\tan(i)=\frac{\alpha}{1+\alpha/\sqrt{3}}, \hspace{0.2in} 
e=\left(1+ \frac{2\alpha}{\sqrt{3}}+\frac{4\alpha^2}{3} \right)^{1/2}-1,
\label{phie}
\end{equation}
and the constellation of spacecraft will fly almost 
like three vertices of a moving equilateral triangle (the distance between any two of 
the spacecraft is not strictly constant) with its center moving along the circular reference orbit. 
Given $\alpha$ and $a$, then, $e$ 
and $i$ can be calculated. Using (\ref{orbit1}), (\ref{orbits}), and (\ref{Ek}), 
the orbits of the three spacecraft can be obtained. The distances between any two of
the three spacecraft are described in Figure \ref{orbitDNKV}, which is very similar to
the one given in \cite{dnkv05}.

However, for the Laser Interferometer Space Antenna (LISA) constellation, it is desired to find 
an optimal solution such that the distances between any two of the spacecraft are as 
close to a desired constant as possible. 

\subsection{NKDV orbit}

\begin{figure}[ht!]
\centering
\includegraphics[width=5in,height=4.5in]{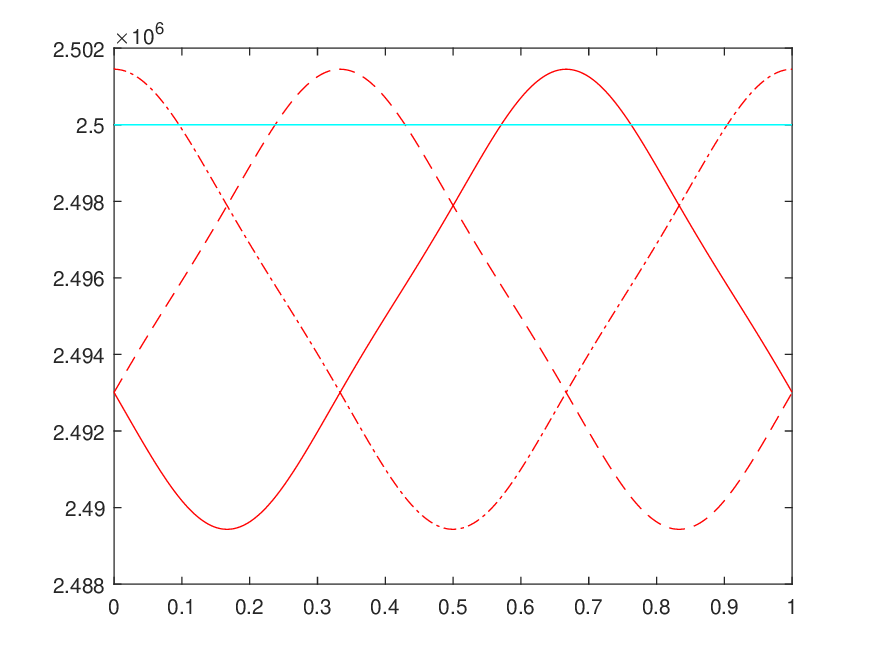}
\caption{Spacecraft distances in one period of NKDV orbit.  The horizontal line is the 
desired distance between any spacecraft pair. The solid lines are the distance variation 
between spacecraft 1 \& 2. The dashed lines are the distance variation between spacecraft 
1 \& 3. The dashed-dot lines are the distance variation between spacecraft 2 \& 3.} 
\label{orbitNKDV}
\end{figure}

To reduce the distance variation, Nayak et al \cite{nkdv06} allows the angle 
$\psi$ to be adjusted around $\pi/3$, i.e., $\psi=\pi/3+\delta$. Using a second-order
approximated model, they found the 
best solution is $\delta=\frac{5}{8}\alpha$, and the following relations hold.
\begin{equation}
\tan(i)=\frac{2}{\sqrt{3}}\frac{\alpha\sin(\pi/3+\delta)}{[1+\frac{2}{\sqrt{3}}\alpha\cos(\pi/3+\delta)]}, 
\hspace{0.2in} 
e=\left(1+ \frac{4\alpha^2}{3}+\frac{4\alpha}{\sqrt{3}} \cos(\pi/3+\delta) \right)^{1/2}-1.
\label{Ephi}
\end{equation}
%Again, $a=R(1+e)$. 
Given $\alpha$, $\delta$, and $a$,  then $e$ and $i$ can be calculated. 
%Applying these eccentricity and inclination to the second-order approximated model, 
%Nayak et al obtained some formulas \cite[(10) and (11)]{nkdv06} to calculate the 
%approximated distance between the spacecraft at any time $t$. Please note that this 
%distance calculation is based on the second-order approximated model. Using these formulas, 
%we can repeat the figure of these approximated distances reported by Martens and Joffre 
%\cite[Figure 3(c)]{mj21} (green lines in Figure \ref{orbitNKDV}). As we explained, 
%the eccentricity and inclination optimized in \cite{nkdv06,mj21} are based 
%on the approximated second-order model and 
%the distances calculated in \cite{nkdv06,mj21} also use the approximated 
%second-order model. What we really need to know is the accurate distances between the 
%spacecraft when the eccentricity and inclination are {\it applied to the accurate nonlinear model}.
Using (\ref{orbit1}), (\ref{orbits}), and (\ref{Ek}), the accurate orbits of the three spacecraft 
can be obtained. The distances between any two of the three spacecraft using accurate
nonlinear model are presented in Figure \ref{orbitNKDV}. It is worthwhile to note that the 
accurate peak-to-peak distance using the nonlinear model for NKDV design is about 12,000 km
which is a significant improvement over DNKV design (28,820  km). But the mean orbits of
the three spacecraft are deviated from the designed $2.5 \times 10^{6}$ km, which is similar
to the reported result in \cite[Figure 3]{mj21}.

\section{The optimal orbit design}

The optimal orbit design is based on the arc-search techniques proposed in
\cite{yang20} for interior-point method.

\subsection{An arc-search infeasible interior-point algorithm}

\begin{figure}[ht!]
\centering
\includegraphics[width=5in,height=3in]{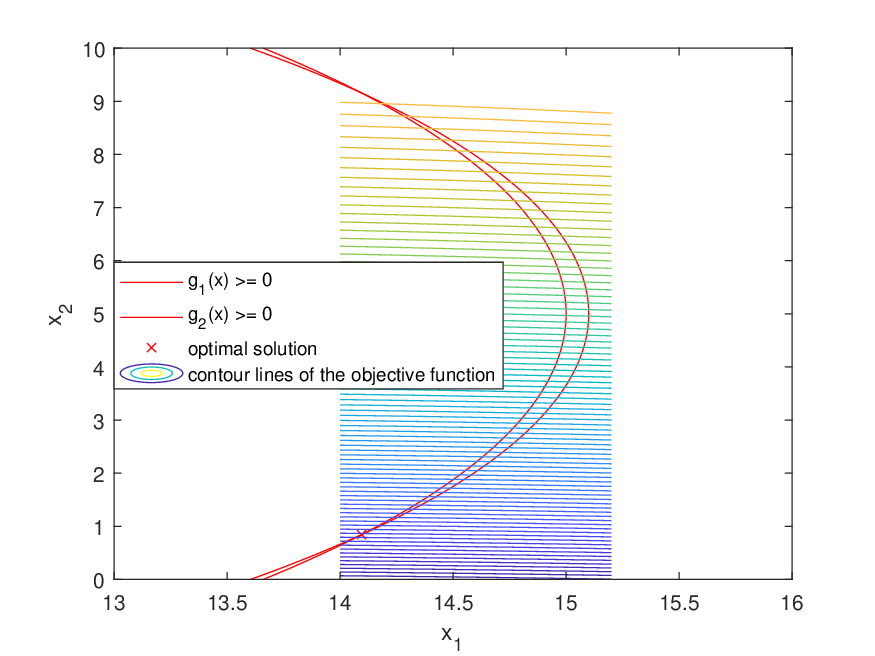}
\caption{The constraint of HS-19.} 
\label{HS19}
\end{figure}

\begin{figure}[ht!]
\centering
\includegraphics[width=5in,height=3in]{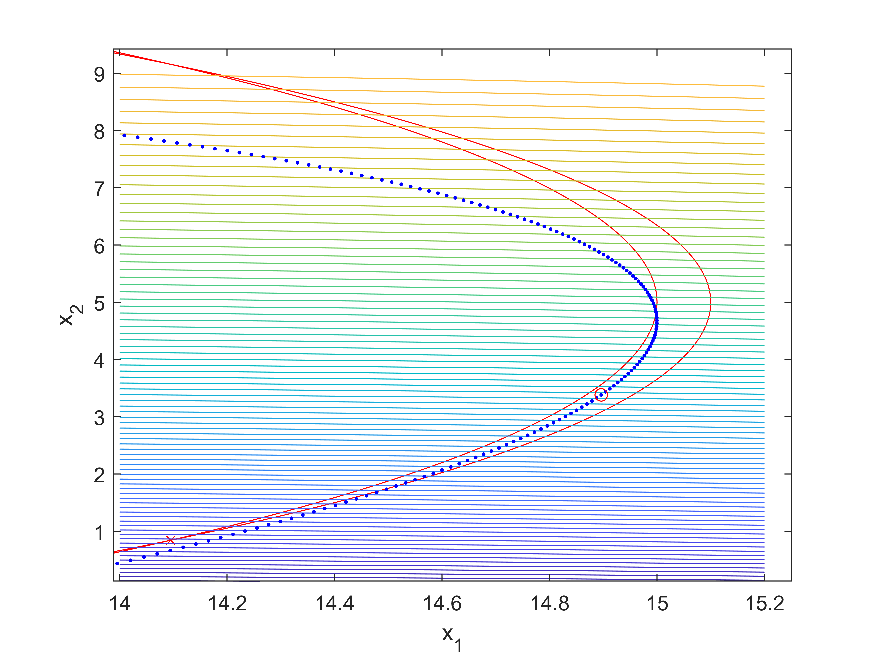}
\caption{The arc used for searching optimizer at current iterate of Problem (\ref{simpleEx}).} 
\label{arcApp}
\end{figure}

Consider a general nonlinear optimization problem written as the following form:
\begin{align}
\begin{array}{rcl}
\min &:& f(\x) \\
\textrm{s.t.} &:& \h(\x) = \0, \label{NP}\\
& & \g(\x) \ge \0,
\end{array}
\end{align}
where $f: \R^n \rightarrow \R$ is the nonlinear objective function, $\h(\x) = \0$ 
represents $m$ the nonlinear equalilty constraints, and $\g(\x) \ge \0$ represents 
$p$ the inequality constraints. The infeasible interior-point method has become 
popular to solve the nonlinear optimization problem \cite{wb05}. The key idea of 
the method is to start with an initial point $\x^0$ that meets the nonlinear inequality 
constraints ($\x^0$ is an interior-point) but may NOT meet the equality constraints 
($\x^0$ is an infeasible point) because it is very expensive to find a solution of the
nonlinear system of equations $\h(\x) = \0$. However, as the iterate sequence 
$\{ \x^k \}$ approximate the optimal solution $\x^*$, the iterates $\x^k$ 
will approach to a feasible solution (meeting both equality and inequality).
The traditional optimization method uses linear search to find a better iterate in
every iteration. This may not be a good strategy because the constraints are 
nonlinear. A benchmark test problem (HS-19) in Hock and Schittkowski \cite{hs81}
test set is used to justify why arc-search is a more appropriate search method.

\begin{align}
\begin{array}{rcl}
\min &:& f(\x)=(x_1-10)^3+(x_2-20)^3 \\
\textrm{s.t.} &:& (x_1-5)^2+(x_2-5)^2-100, \ge 0 \\
& & -(x_2-5)^2 - (x_1-6)^2+82.81 \ge 0,  \\
& & 13 \le x_1 \le 100, \\
& & 0 \le x_2 \le 100.
\end{array}
\label{simpleEx}
\end{align}
For this problem, the last 4 boundary inequalities are redundent. The area of the constraint
is depicted in Figure \ref{HS19}, which is between two red curves. The contour lines 
represent the levels of the objective function, which decrease in the top-down direction. 
The optimal solution is at the interception of the two red curves marked with a red 'x'. 
Clearly, for an iterate inside the area of the constraint, searching along a 
decreasing line segment for optimizer is not as good as searching along an arc
described in Figure \ref{arcApp} which is part of an ellipse.

For problem (\ref{simpleEx}) depicted in Figure \ref{arcApp}, 
the arc (represented in blue color) passing the current iterate (marked with a red 'o')
can be obtained systematically and be used to search for the optimizer. The merit 
of using arc-search for the interior-point method  can be seen from Figure \ref{arcApp} 
and was analyzed in an internal report \cite{yph23,yang25} where an efficient algorithms is 
proposed and the convergence is analyzed\footnote{The algorithm is an improved 
version of the arc-search infeasible interior-point algorithm of \cite{yiy22}}.
An optimization tool that implements the algorithm in \cite{yang25} by the author and
provides the ability to compute first and second order 
derivatives via an automatic differentiation\footnote{Details about automatic
differentiation are discussed in \cite[Section 7.2]{nw06}.} tool of \cite{rump23}
with improved the speed and accuracy was used to solve the optimal orbit design
problem (\ref{optPro}) to be discussed.

\begin{figure}[ht!]
\centering
\includegraphics[width=5in,height=4.7in]{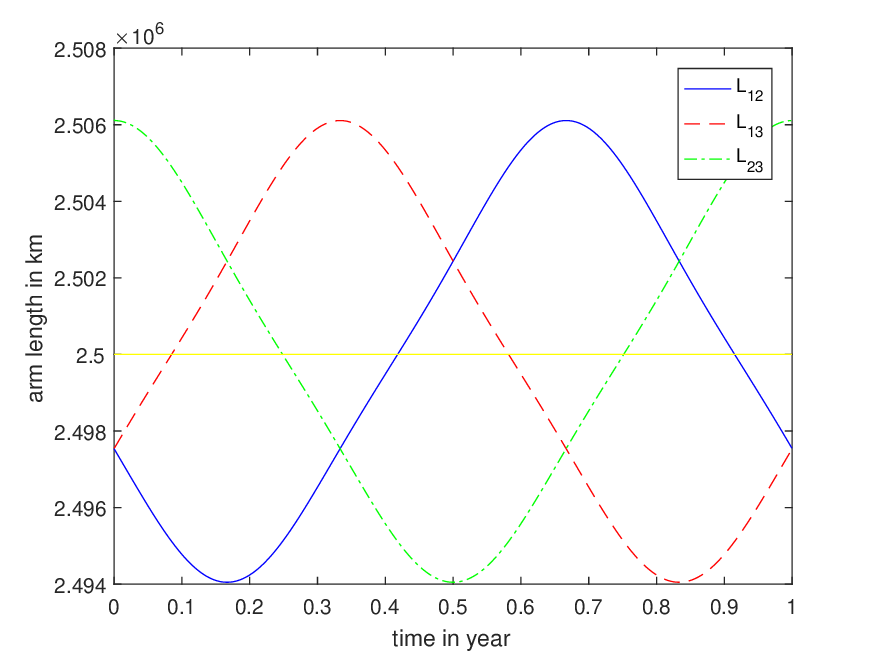}
\caption{Spacecraft distances in one period of optimal orbit.  The horizontal line is the desired distance
between any spacecraft pair. The solid line is the distance variation between spacecraft 1 \& 2.
The dashed line is the distance variation between spacecraft 1 \& 3. The dashed-dot line is the 
distance variation between spacecraft 2 \& 3.} 
\label{optOrbit}
\end{figure}

\subsection{Optimal orbit design assuming identical $e_k$ and $i_k$ for all spacecraft $k$}\label{main}

\begin{figure}[ht!]
\centering
\includegraphics[width=5in,height=4.7in]{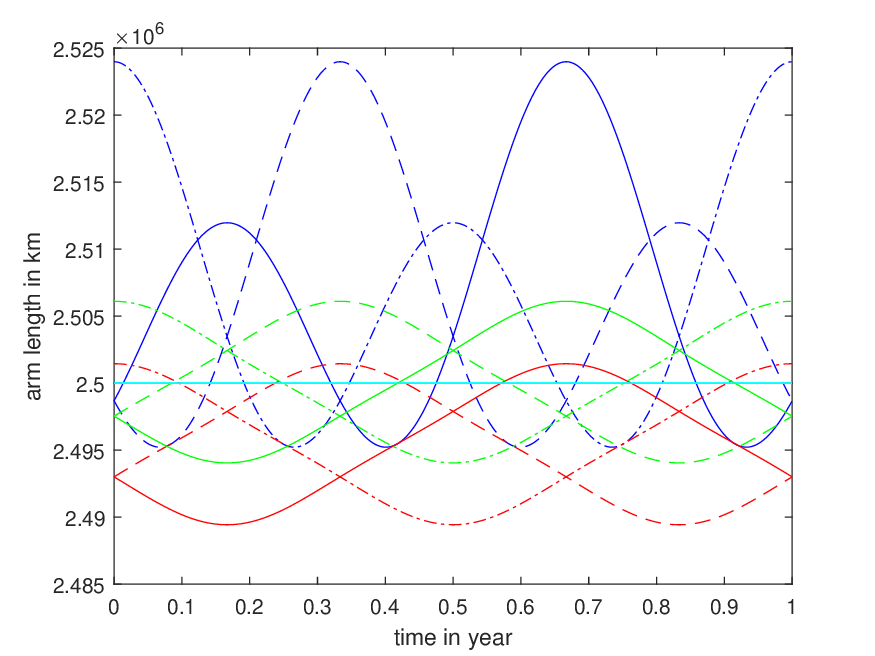}
\caption{Distances comparison of all three orbit designs. The capri line is the desired distance
between any spacecraft pair. The blue lines are orbit distances between spacecraft pairs
of DNKV orbit design. The red lines are orbit distances between spacecraft pairs
of NKDV orbit design. The green lines are orbit distances between spacecraft pairs
of the optimal orbit design. The solid lines are orbit distances between spacecraft 1 \& 2.
The dashed lines are orbit distances between spacecraft 1 \& 3. The dashed dot 
lines are orbit distances between spacecraft 2 \& 3.} 
\label{compare}
\end{figure}

Without loss of generality, we may assume each spacecraft's orbit is elliptic meaning its
size and shape depend only on $a$ and $e$. Since the constellation's period should be the same
as the Earth's period, this means that $a=R$. Each spacecraft orbit is inclined with respect to the
ecliptic plan by an angle $i$. Therefore, only $e$ and $i$ are independent, 
and should be optimized ($\delta$ is implicitly determined by $i$ therefore is redundant). 
We formalize the orbit design problem as an optimization problem that directly selects 
the optimal $i$ and $e$ to minimize the distance deviation from a desired constant, 
while meeting all constraints defined by (\ref{orbit1}),  (\ref{orbits}), and (\ref{Ek}). 

Let $T=\frac{2\pi}{\Omega}$ be the period of the three spacecraft orbits.
For $0 = t_0 < t_1 < \ldots <t_{n-1} <t_n =  T$, the desired distance between
spacecraft $i$ and $j$ (for $1 \le i < j \le 3$) is 
$\ell_{i,j}= 2,500,000 \hspace{0.1in} \mbox{km}$, and the distance deviation 
between spacecraft $i$ and $j$ at $t_k$ from the desired distance is given by
\begin{equation}
d_{i,j}(t_k)=\sqrt{(X_i(t_k)-X_j(t_k))^2+(Y_i(t_k)-Y_j(t_k))^2+(Z_i(t_k)-Z_j(t_k))^2}-\ell_{i,j}.
\label{obj}
\end{equation}

Our objective is to minimize the accumulative distance deviation (\ref{obj}) for all $t_k$, 
subject to satisfying the orbital requirements of (\ref{orbit1}), (\ref{orbits}), and (\ref{Ek}) 
for all $t_k$. In addition, we impose some safe boundary constraints on $e$ and $i$, 
aiming at accelerating the convergence rate. Therefore,
the optimization problem can be written as follows.
\begin{align}
\begin{array}{rcl}
\min &  &  \sum_{t_k=0,\ldots,T} \sum_{1 \le i < j \le 3} d_{i,j}^2(t_k) \\
\textrm{s.t.} & & E_1(t_k) + e \sin(E_1(t_k)) = \Omega t_k , \\
& & E_2(t_k) + e \sin(E_2(t_k)) = \Omega t_k - \frac{2\pi}{3}, \\
& & E_3(t_k) + e \sin(E_3(t_k)) = \Omega t_k - \frac{4\pi}{3}, \\
& & X_1(E_1(t_k))=a (\cos(E_1(t_k)) + e) \cos(i), \\ 
& & Y_1(E_1(t_k))=a \sqrt{1-e^2} \sin(E_1(t_k)), \\ 
& & Z_1(E_1(t_k))=a (\cos(E_1(t_k)) + e) \sin(i)  \\
& & X_2(t_k)=X_1(E_2(t_k)) \cos\left(\frac{2\pi}{3} \right)- Y_1(E_2(t_k)) \sin \left( \frac{2\pi}{3}  \right), \\
& & Y_2(t_k)=X_1(E_2(t_k)) \sin\left(\frac{2\pi}{3} \right) + Y_1(E_2(t_k)) \cos \left( \frac{2\pi}{3} \right), \\
& & Z_2(t_k)=Z_1(E_2(t_k)), \\
& & X_3(t_k)=X_1(E_3(t_k)) \cos\left(\frac{4\pi}{3} \right)- Y_1(E_3(t_k)) \sin \left( \frac{4\pi}{3} \right), \\
& & Y_3(t_k)=X_1(E_3(t_k)) \sin\left(\frac{4\pi}{3} \right) + Y_1(E_3(t_k)) \cos \left( \frac{4\pi}{3} \right), \\
& & Z_3(t_k)=Z_1(E_3(t_k)), \\
& & 0 = {\Omega}t_0 <{\Omega} t_1 < \ldots <{\Omega}t_{n-1} <{\Omega}t_n =  2\pi, \\
& &  0 \le e \le 0.01, \hspace{0.1in} 0 \le i \le \pi/6,
\label{optPro}
\end{array}
\end{align}
where $i$ and $e$ are independent variables, and $0 \le \Omega t_k \le 2\pi$. 
For the fixed $i$ and $e$, it is clear that $E_1$, $E_2$, and $E_3$ are functions 
of $t_k$. In addtion, $X_i$, $Y_i$, and $Z_i$ are functions of $E_i(t)$, $i=1, 2, 3$.

Starting from a feasible solution $(e,i)=(0.0047975,0.008315)$ (which is determined
based on some trial-and-error process), after $14$ iterations, we find the optimal solution 
%$$(e^*,i^*)=(0.004275864892411,0.007405911848104).$$
$$(e^*,i^*)=(0.004824385965325,0.008355663130457).$$
The distances between any two of the three spacecraft are obtained from 
(\ref{orbit1}),  (\ref{orbits}), and (\ref{Ek}). The result is presented in Figure \ref{optOrbit}.
The comparison of all three designs is presented in Figure \ref{compare}.
The distance change of DNKV design is about $2.8789e+04$ kilometers,
the distance change of NKDV design is about $1.2e+04$ kilometers, which is essentially
the same one as the optimal design. But latter is centered about $2.5 \times 10^6$ km
while the former is not.

\subsection{Optimal orbit design assuming different $e_k$ and $i_k$ for all spacecraft $k$}

In this design, we assume that the eccentricity and inclination of spacecraft $k$ are
$e_k$ and $i_k$ for $k=1,2,3$. We would like to know, under this assumption, if we
can find a better optimal design using the extra degrees of freedom. For this purpose, 
Kepler's equation for the three spacecraft is given as follows.
\begin{equation}
E_k + e_k \sin(E_k) = \Omega t - (k-1)\frac{2\pi}{3},
\hspace{0.1in} \mbox{for $k=1,2,3$}.
\label{Ekek}
\end{equation}
Accordingly, the orbits of the three spacecraft at the same orientation are modified as
\begin{subequations}
\begin{gather}
\tilde{X}_k(E_k)=a (\cos(E_k) + e_k) \cos(i_k), \mbox{for $k=1,2,3$,}  \\ 
\tilde{Y}_k(E_k)=a \sqrt{1-e_k^2} \sin(E_k), \hspace{0.1in} \mbox{for $k=1,2,3$,} \\ 
\tilde{Z}_k(E_k)=a (\cos(E_k) + e_k) \sin(i_k), \mbox{for $k=1,2,3$} .
\end{gather}
\label{orbitk}
\end{subequations}
Finally, the desired orbits of the three spacecraft are given by
\begin{subequations}
\begin{gather}
X_k=\tilde{X}_k(E_k) \cos\left(\frac{2\pi}{3} (k-1) \right)- \tilde{Y}_k(E_k) 
\sin \left( \frac{2\pi}{3} (k-1)  \right), \hspace{0.1in} \mbox{for $k=1,2,3$,}  \\
Y_k=\tilde{X}_k(E_k) \sin\left(\frac{2\pi}{3} (k-1) \right) + 
\tilde{Y}_k(E_k)\cos \left( \frac{2\pi}{3} (k-1)  \right),  
\hspace{0.1in} \mbox{for $k=1,2,3$,} \\
Z_k=\tilde{Z}_k(E_k), \hspace{0.1in} \mbox{for $k=1,2,3$} .  
\end{gather}
\label{orbitsK}
\end{subequations}

Combining all above formulas yields the optimization problem:

\begin{align}
\begin{array}{rcl}
\min &  &  \sum_{t_k=0,\ldots,T} \sum_{1 \le i < j \le 3} d_{i,j}^2(t_k) \\
\textrm{s.t.} & & E_1(t_k) + e_1 \sin(E_1(t_k)) = \Omega t_k , \\
& & E_2(t_k) + e_2 \sin(E_2(t_k)) = \Omega t_k - \frac{2\pi}{3}, \\
& & E_3(t_k) + e_3 \sin(E_3(t_k)) = \Omega t_k - \frac{4\pi}{3}, \\
& & X_1(E_1(t_k))=a (\cos(E_1(t_k)) + e_1) \cos(i_1), \\ 
& & Y_1(E_1(t_k))=a \sqrt{1-e_1^2} \sin(E_1(t_k)), \\ 
& & Z_1(E_1(t_k))=a (\cos(E_1(t_k)) + e_1) \sin(i_1)  \\
& & \tilde{X}_2(E_2(t_k))=a (\cos(E_2(t_k)) + e_2) \cos(i_2), \\
& & \tilde{Y}_2(E_2(t_k))=a \sqrt{1-e_2^2} \sin(E_2(t_k)), \\ 
& & \tilde{Z}_2(E_2(t_k))=a (\cos(E_2(t_k)) + e_2) \sin(i_2)  \\
& & \tilde{X}_3(E_3(t_k))=a (\cos(E_3(t_k)) + e_3) \cos(i_3), \\
& & \tilde{Y}_3(E_3(t_k))=a \sqrt{1-e_3^2} \sin(E_3(t_k)), \\ 
& & \tilde{Z}_3(E_3(t_k))=a (\cos(E_3(t_k)) + e_3) \sin(i_3)  \\
& & X_2=\tilde{X}_2(E_2) \cos\left(\frac{2\pi}{3} \right)- \tilde{Y}_2(E_2) 
\sin \left( \frac{2\pi}{3}  \right), \\
& & Y_2=\tilde{X}_2(E_2) \sin\left(\frac{2\pi}{3} \right) + 
\tilde{Y}_k(E_k)\cos \left( \frac{2\pi}{3}   \right), \\  
& & Z_2=\tilde{Z}_2(E_2), \\
& & X_3=\tilde{X}_3(E_3) \cos\left(\frac{4\pi}{3} \right)- \tilde{Y}_3(E_3) 
\sin \left( \frac{4\pi}{3}  \right), \\
& & Y_3=\tilde{X}_3(E_3) \sin\left(\frac{4\pi}{3} \right) + 
\tilde{Y}_k(E_3)\cos \left( \frac{4\pi}{3}   \right),  \\
& & Z_3=\tilde{Z}_3(E_3), \\
& & 0 = {\Omega}t_0 <{\Omega} t_1 < \ldots <{\Omega}t_{n-1} <{\Omega}t_n =  2\pi, \\
& &  0 \le e \le 0.01, \hspace{0.1in} 0 \le i \le \pi/6,
\label{optPro1}
\end{array}
\end{align}

Starting from the following feasible point 
$$(e_1,i_1,e_2,i_2,e_3,i_3,)=(0.0047975,0.008315,0.0047975,0.008315,0.0047975,0.008315),$$
after $14$ iteration, we obtain an optimal solution
$$(e_1,i_1,e_2,i_2,e_3,i_3,)=(0.0048244,0.0083556,0.0048243,0.0083556,0.0048243,0.0083556),$$
which is essentially the same result we obtained in the previous section. Starting from some
random initial points near the above optimal solution, we reached the same result. This means
that the optimal solution of (\ref{optPro}) is likely a global optimal solution of (\ref{optPro1}).
Since problem (\ref{optPro}) is simpler than problem (\ref{optPro1}), it is more efficient to
solve (\ref{optPro}) than solve (\ref{optPro1}).

\subsection{Extension to $n$ spacecraft in formation fly}

We have discussed the solution of $3$ spacecraft in formation fly. The method discussed in Section
\ref{main} can easily be extened to the case of $n \ge 4$ spacecraft in formation fly. The key idea is to use
(\ref{kepler}) and (\ref{orbit1}) to represent the orbit of Spacecraft 1. Then the orbit presentation for 
the $k$-th ($k=2,\ldots, n$) spacecraft can be obtained by rotating the orbit of Spacecraft 1 by 
$360(k-1)/n$ degrees about the center of the coordinate system (the Sun, see Figure \ref{orbitGeneral}).
Following exactly the same procedure in Section \ref{main}, we can obtain an optimization problem
similar to (\ref{optPro}), and the optimization algorithm/tool developed in \cite{yang25} 
can be used to solve the general problem.

\section{Conclusions}\label{sec:conclusion}

The LISA orbit design problem is formulated as a nonlinear optimization problem using 
exact nonlinear Kepler's orbit equations. The problem is solved by 
using an arc-search interior-point algorithm. The solution minimizes the distance
variations about a designed constant between the three LISA spacecraft, thereby
reducing the Doppler shift and breathing angle effects on the measurement of gravitational waves.

\section{Acknowledgements}

This work is supported in part by NASA's IRAD 2023 fund SSMX22023D.
The author thanks Dr. Pritchett at Goddard Space Flight Center of NASA 
for his valuable comments and suggestions 
that helped to improve the presentation of the paper.

\section{Data availability statement}

The matlab code that is used to generate the result is available upon reasonable request.

%
% \section*{Conflict of interest}
%
% The authors declare that they have no conflict of interest.

%\appendix

\end{document}